\documentclass[aps,prb,twocolumn,floatfix,superscriptaddress]{revtex4}
\usepackage{epsf}
\usepackage{graphics}

\newcommand{\myscalebox}[1]{\scalebox{0.35}[0.35]{#1}}

\begin{document}


\title{The stiffness exponent of two-dimensional Ising spin glasses
  for non-periodic boundary conditions using aspect-ratio scaling}
\author{Alexander K. Hartmann}
\email{hartmann@theorie.physik.uni-goettingen.de}
\affiliation{Department of Physics, University of California,
Santa Cruz CA 95064, USA}
\affiliation{Ecole Normale Sup\'erieure,
24, Rue Lhomond,
75231 Paris Cedex 05, France}
\affiliation{Institut f\"ur Theoretische Physik, Universit\"at G\"ottingen,
 Bunsenstr.9, 37073 G\"ottingen, Germany}
\author{Alan J. Bray}
\affiliation{Department of Physics and Astronomy, The University of
  Manchester, Manchester M13 9PL, UK}
\author{A. C. Carter}
\affiliation{Department of Physics and Astronomy, The University of
  Manchester, Manchester M13 9PL, UK}
\author{M. A. Moore}
\affiliation{Department of Physics and Astronomy, The University of
  Manchester, Manchester M13 9PL, UK}
\author{A. P. Young}
\affiliation{Department of Physics, University of California,
Santa Cruz CA 95064, USA}
\affiliation{Theoretical Physics, 1, Keble Road, Oxford OX1 3NP, UK}
\date{\today}

\begin{abstract}
We study the scaling behavior of domain-wall energies in
two-dimensional Ising spin glasses with Gaussian and bimodal
distributions of the interactions and different types of boundary 
conditions. The domain walls are generated by changing the boundary
conditions at $T=0$ in one direction. The ground states of the 
original and perturbed system are calculated numerically by applying 
an efficient matching algorithm. Systems of size $L\times M$ with 
different aspect-ratios $1/8\le L/M\le 64$ are considered. 
For Gaussian interactions, using the {\em aspect-ratio scaling} 
approach, we find a stiffness exponent $\theta=-0.287(4)$, which 
is independent of the boundary conditions in contrast to earlier 
results. Furthermore, we find a scaling behavior of the
domain-wall energy as predicted by the aspect-ratio approach. Finally,
we show that this approach does not work for the bimodal distribution
of interactions.

\end{abstract}

\maketitle

\section{Introduction}
Although spin glasses\cite{reviewSG} have been studied for more 
than two decades, finite-dimensional spin glasses
are still far from being well understood. This is true 
also in $d=2$ dimensions, where no stable spin-glass
phase\cite{mcmillan1984,bray84,rieger1996,stiff2d} exists at $T>0$.
The fact that the transition temperature, $T_c$, is zero
can be seen by studying domain walls between
ground states induced by changing the boundary conditions. For a
system of size $L\times M$, if the domain wall is forced to run in the
$y$-direction, the mean domain-wall energy scales like $M^{\theta}$ in
the thermodynamic limit, where $\theta$ is the {\em stiffness exponent}. 
If $\theta<0$, as in two-dimensional spin glasses, no stable phase can 
exist at finite temperature. 

Recently\cite{stiff2d} slightly different values of the stiffness 
exponent have been found for Gaussian distribution of the bonds 
for different choices of the boundary conditions. Here, by
applying the aspect-ratio method\cite{carter2002}, we show that 
this difference is due to corrections to scaling which persisted, 
at least for one set of boundary conditions, even for the quite 
large system sizes used in Ref.~\onlinecite{stiff2d}. The results 
presented here are consistent with $\theta=-0.287(4)$ for all boundary 
conditions. This agrees with the result found in
Ref.~\onlinecite{carter2002} where periodic boundary conditions were
applied in all directions for smaller systems than investigated here, 
indicating a high degree of universality
for the exponent $\theta$. Apart from using larger systems and
different boundary conditions, we go in another respect beyond the
previous work, because aspect ratios, $R \equiv
L/M$, less than unity are studied here. 
This enables us to verify the proposed scaling
behavior in the region $R<1$, where the finite-size scaling depends on the
 boundary conditions.

The model we study consists  of $N=L\times M$\ Ising spins 
$S_i=\pm 1$ on a square lattice with the Hamiltonian
\begin{equation}
{\cal H} = -\sum_{\langle i,j \rangle} J_{ij} S_iS_j\,,
\end{equation}
where the sum runs over all pairs of nearest neighbors $\langle i,j \rangle$
and the $J_{ij}$ are quenched random variables. Here, we consider
two kinds of disorder distributions: 
\begin{enumerate}
\item[(i)]
Gaussian with zero mean and variance unity,
\item[(ii)]
bimodal, i.e.\ $J_{ij}=\pm 1$ with equal probability.
\end{enumerate}

To study whether an ordered phase is stable at finite temperatures,
the following procedure is usually applied 
\cite{mcmillan1984,bray84,rieger1996,kawashima1997}.
First a ground state 
of the system is calculated, having energy $E_0$.
Then the system is perturbed to introduce a domain wall and the
new ground state energy, $E_0^{\rm pertb}$ is evaluated.
For each sample, the domain-wall energy is given by 
$E_{\rm DW}=|E_0^{\rm pertb}-E_0|$.
Here, we apply free boundary conditions (bc) in the $y$-direction. To study the
scaling behavior of the domain-wall energy, we consider two means to
introduce domain walls:

\begin{enumerate}
\item[(i)]
``P-AP'': First a ground state with periodic bc in the
$x$-direction is calculated. Then the system is perturbed by
introducing antiperiodic bc in that directions, e.g.\ by reversing 
one line of bonds parallel to the $y$-direction.

\item[(ii)]
``F-DW'' (called ``F-AF'' in Ref.~\onlinecite{carter2002}):
First a ground state with free bc in the
$x$-direction is calculated. For the perturbed system, we 
then add extra bonds which wrap around the system in the
$x$-direction, and have a sign and strength such that they force the spins
they connect to have the opposite relative orientation to that which
they had in the original ground state.
\end{enumerate}

In both cases,
$M$ denotes the size of the edge along which the boundary
conditions are changed to induce a domain wall (i.e.\ the size in the 
$y$-direction).
The 
scaling behavior 
\begin{equation}
\langle E_{\rm DW}\rangle=M^{\theta}F(L/M)\end{equation}
has been
predicted\cite{carter2002}, 
with the following limiting forms:
\begin{equation}
F(R) \sim
\left\{
\begin{array}{lll}
R^{-1} &  & \mbox{for $R \to \infty$} \\
R^{\theta-(d-1)/2} & \mbox{(P-AP)} & \mbox{for $R \to 0$} \\
R^{\theta-(d-1)}   & \mbox{(F-DW)} & \mbox{for $R \to 0$} .
\end{array}
\right.
\label{limiting_forms}
\end{equation}
We shall verfiy that the data does follow this scaling form very well.

\section{The Algorithm}
In greater than two dimensions, or in the presence of a magnetic field,
the exact calculation of spin-glass ground states belongs
to the class of NP-hard problems\cite{barahona1982,opt-phys2001}. 
This means that only algorithms with exponentially increasing running 
times are known. However, for the special case of a planar system without 
magnetic field, e.g. a square lattice with periodic boundary conditions 
in at most one direction, there are efficient polynomial-time ``matching'' 
algorithms\cite{bieche1980}. The basic idea is to represent each realization 
of the disorder by its frustrated plaquettes\cite{toulouse1977}. Pairs of
frustrated plaquettes are connected by paths in the lattice and the weight 
of a path is defined by the sum of the absolute values of the coupling 
constants which are crossed by the path. A ground state corresponds to the 
set of paths with minimum total weight, such that each frustrated plaquette 
is connected to exactly one other frustrated plaquette. This is called a 
minimum-weight perfect matching. The bonds which are crossed by paths
connecting the frustrated plaquettes are unsatisfied in the ground state, 
and all other bonds are satisfied. 

For the calculation of the minimum-weight perfect matching, efficient
polynomial-time algorithms are available\cite{barahona1982b,derigs1991}.
Recently, an implementation has been presented\cite{palmer1999}, where 
ground-state energies of large systems of size $N\le 1800^2$ were calculated. 
Here, an algorithm from the LEDA library\cite{leda1999} has been applied,
which limits the system sizes due to the restricted size of the main memory 
of the computers which we used. Furthermore, each system border with a free 
boundary decreases the running time relative to a boundary with pbc. 
The reason is that frustrated plaquettes near a free boundary may
be connected even when they are far apart from each other, increasing the
number of possible connections. On the other hand, full pcb cannot be
realized with the matching algorithm. 
Hence, we have limited our system sizes to $L\le 1024, M\le 16$
for $R\ge 1$, and $L\le 12, M\le 768$ for $R< 1$.

\section{Results}
For Gaussian interactions we have considered systems with P-AP and 
F-DW boundary conditions with aspect ratios $R=1/8$ to $R=64$. We took
system sizes $L=1$ to $L=12$ for $R=1/8$, $1/4$, $1/2$, and
$M=1$ to $M=16$ for $R=1,2,4,\dots,64$. 
Typically $\sim 10^5$ samples were treated
per system size $(L,M)$, except for few combinations, notably 
$R=64$ where we studied $\sim 10^4$ samples.

For bimodal interactions we have studied P-AP boundary conditions 
for $M=1$ to $M=16$ ($R=1,2,4,\dots,64$), with $\sim 10^4$ samples
per system size.

\begin{figure}[htb]
\begin{center}
\myscalebox{\includegraphics{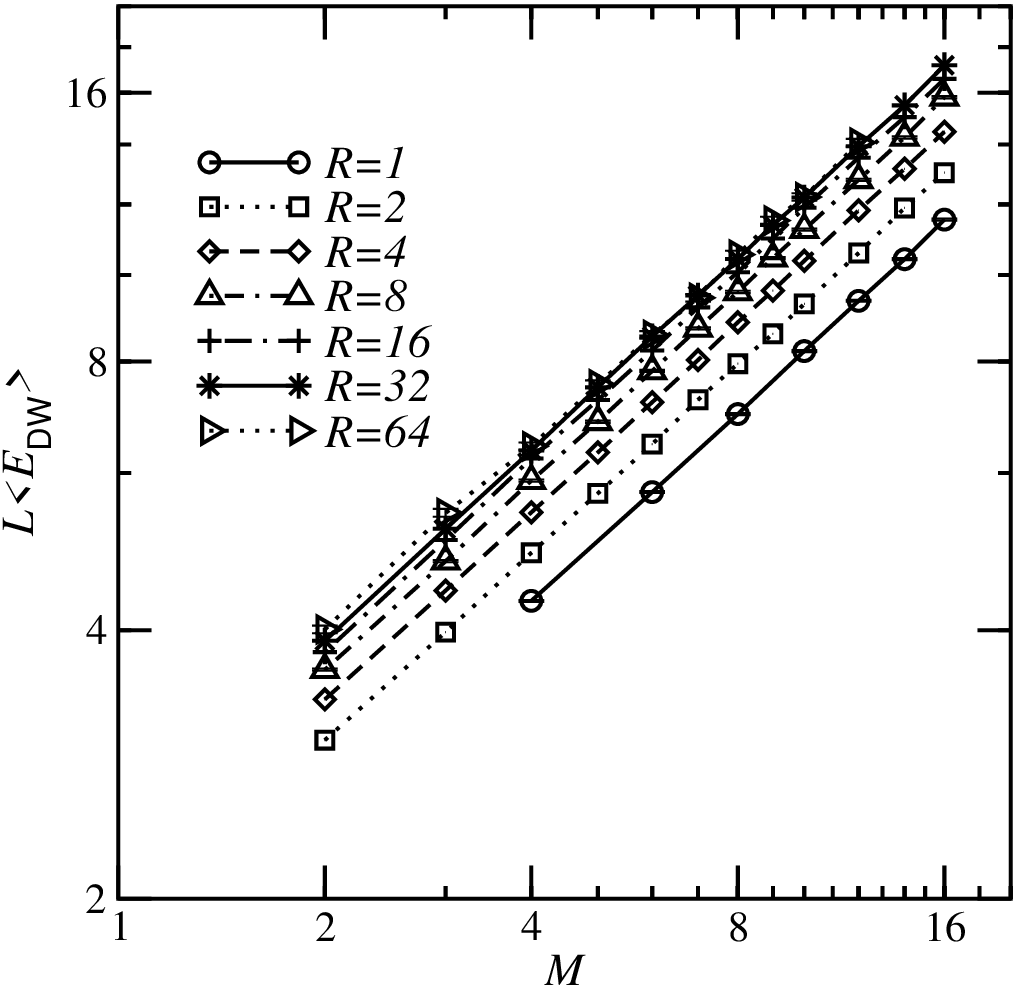}}
\end{center}
\caption{Average domain-wall energy $\langle E_{\rm DW}\rangle$ of Gaussian
system with P-AP boundary conditions as function
  of width $M$ for different aspect ratios $R$. Lines are guides to
  the eyes only.}
\label{figGaussf0Stiff}
\end{figure}

\begin{figure}[htb]
\begin{center}
\myscalebox{\includegraphics{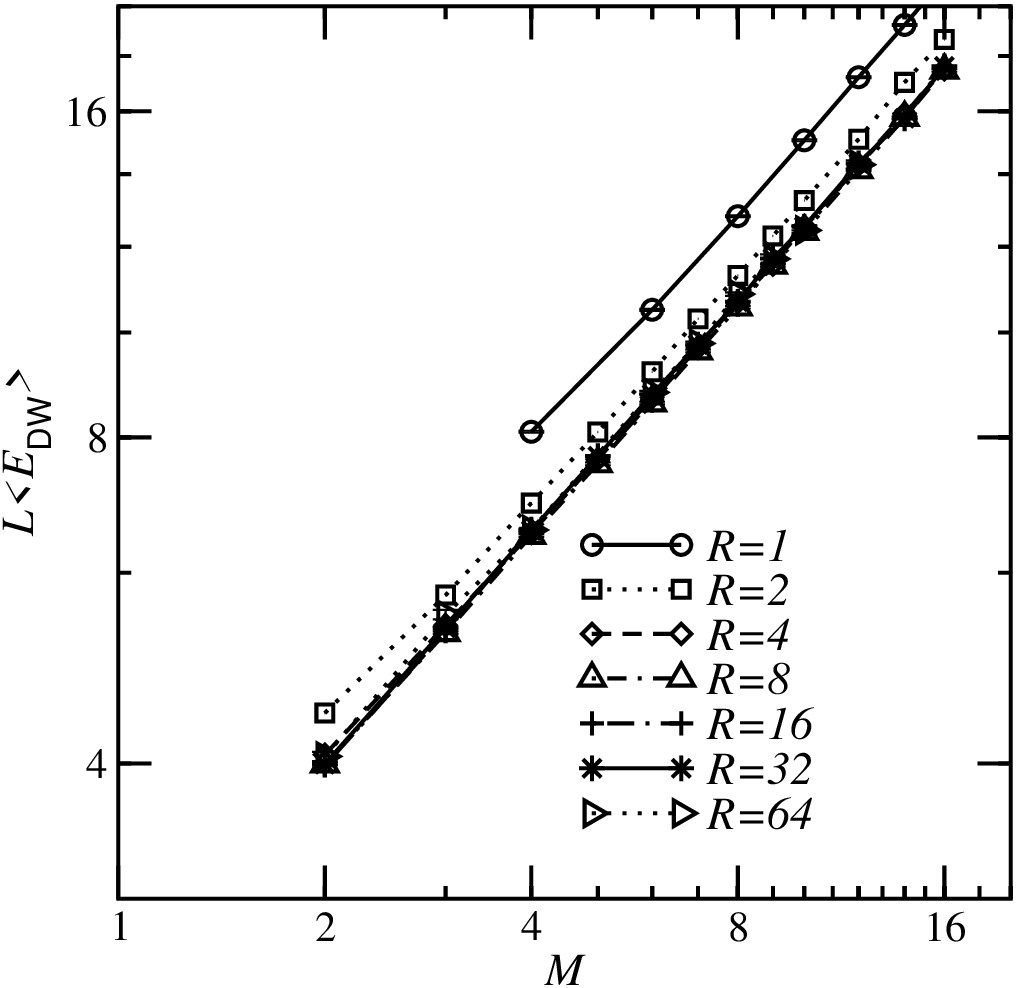}}
\end{center}
\caption{Average domain-wall energy $\langle E_{\rm DW}\rangle$ of Gaussian
  system with F-DW  boundary conditions as function
  of width $M$ for different aspect ratios $R$. Lines are guides to
  the eyes only.}
\label{figGaussf1Stiff}
\end{figure}

We start with the results for the Gaussian distribution. In Figs.\ 
\ref{figGaussf0Stiff} and \ref{figGaussf1Stiff} the scaled
average domain-wall energies $L\langle E_{\rm DW}\rangle$ are displayed
for P-AP and F-DW boundary conditions respectively. From the scaling
theory\cite{carter2002} a $M^{1+\theta}$ behavior is expected for
$R\gg 1$. Indeed
for larger values of $R$ and $M$ straight lines are visible in the
double logarithmic plot. From fitting to algebraic functions, we have
obtained for both cases the values of $\theta$ as a function of $R$;
for small aspect-ratio the small system sizes were omitted from the
fits. The results are displayed in Table~\ref{theta_vals}.

\begin{table}[ht]
\begin{tabular}{|l|c|c|c|}
\hline
$R$ & $\theta_{\rm P-AP}$ & $\theta_{\rm F-DW}$ & 
$\theta^{\rm P}_{\rm F-DW}$ \\
\hline\hline
1 & -0.285(3) & -0.271(2) & -0.153(2) \\
2 & -0.288(3) & -0.279(4) & -0.215(2) \\
4 & -0.288(1) & -0.286(2) & -0.249(2) \\
8 & -0.284(2) & -0.282(2) & -0.265(2) \\
16 &  -0.288(2) & -0.284(3) &-0.273(2) \\
32 & -0.289(2) & -0.289(3) &-0.274(3) \\
64 & -0.290(7) & -0.288(7) &   \\
\hline
\end{tabular}
\caption{Resulting stiffness exponents from fits to functions
  $L\langle E_{DW} \rangle =aM^{1+\theta}$ for P-AP and F-DW boundary 
conditions. The final column gives the result for F-DW boundary conditions  
when pbc are applied in the $y$-direction 
(data from Ref.~\onlinecite{carter2002}).}
\label{theta_vals}
\end{table}

For the P-AP case, the exponent $\theta$ is more or less independent
of the system size. By contrast, for the F-DW case
a significant increase with
system size is observed, with the value seeming to converge near the
value of the P-AP case for large $R$.
This $R$-dependence explains why in previous
work\cite{stiff2d} ($R=1$) a difference between P-AP and F-DW has
been found. The result found here is compatible with the exponents
being equal for both cases. We quote $\theta=-0.287(4)$, which is a
conservative estimate including all values of $\theta$ found for the
P-AP case. This is compatible with the result of $\theta=-0.282(3)$
obtained\cite{carter2002} for systems with full periodic bc.

Part of 
the data in Ref.~\onlinecite{carter2002} corresponds to the F-DW case 
with pbc in the $y$-direction, and we show this here in the 
final column in Table~\ref{theta_vals}, where
$M \le 12$ was used.
For this case an even stronger dependence of the effective 
exponent on $R$ is observed, indicating that corrections to scaling 
are even larger in this case. We suggest that this is because the pbc 
forces the domain wall to take at the top and bottom edges 
the same positions along the
$x$-axis, thereby raising its 
typical energy. In Ref.~\onlinecite{carter2002} it was shown  
that $\theta^{\rm P}_{\rm F-DW}$ (called $\theta_{\rm F-AF}$ in  
Ref.~\onlinecite{carter2002}) nevertheless extrapolates, for 
$R \to \infty$, to an asymptotic value consistent with that obtained 
from the present work, where we employed free boundary conditions
in the $y$-direction.

\begin{figure}[htb]
\begin{center}
\myscalebox{\includegraphics{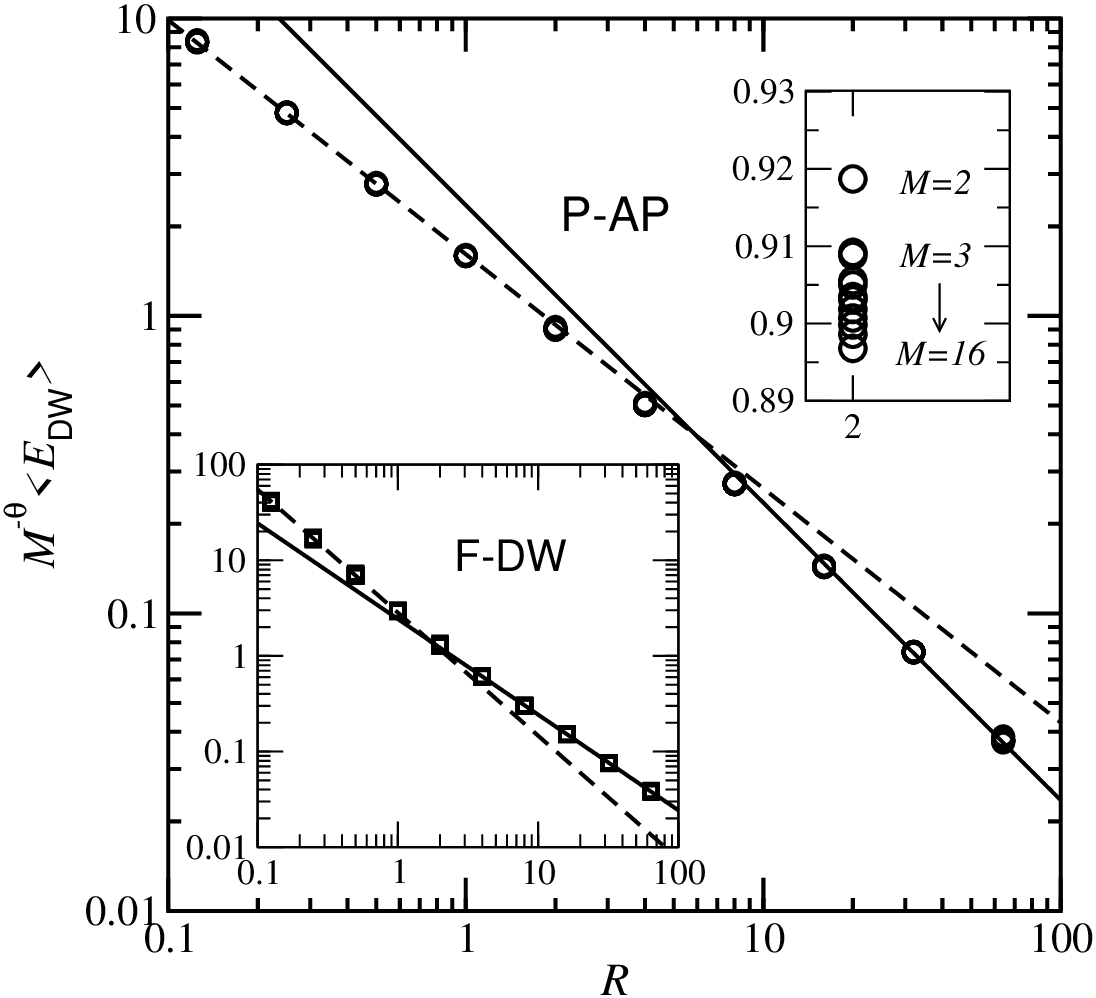}}
\end{center}
\caption{Scaling plot  of $M^{-\theta}\langle E_{\rm DW}\rangle$,
with $\theta=-0.287$, as a
  function of $R$ for P-AP (main plot) and F-DW (left inset) boundary
  conditions for all values of $M$ which have been considered. The
  curve is indeed a (very good) collapse of the results for different
  values of $M$, as exemplified for the P-AP case by the right inset, 
which shows a blow-up of the $R=2$ points.  
The lines represent functions $\sim R^{-1}$ (continuous lines), $\sim
  R^{\theta-(d-1)/2}$ (dashed line, main plot), and
$\sim R^{\theta-(d-1)}$ (dashed line, left inset).}
\label{figScaleR}
\end{figure}

We have also checked the scaling predictions\cite{carter2002}
for $\langle E_{\rm DW}\rangle$ explicitly. In Fig.\ \ref{figScaleR} 
the scaled domain-wall energy $M^{-\theta}\langle E_{\rm DW}\rangle$ is
plotted for all considered values of $M$ as a function of the aspect
ratio $R$. A very good data collapse can be observed (see e.g. right
inset of Fig.\ \ref{figScaleR} for P-AP). 
The functional form of the scaling function agrees very well
with the predictions shown in Eq.~(\ref{limiting_forms}).

\begin{figure}[htb]
\begin{center}
\myscalebox{\includegraphics{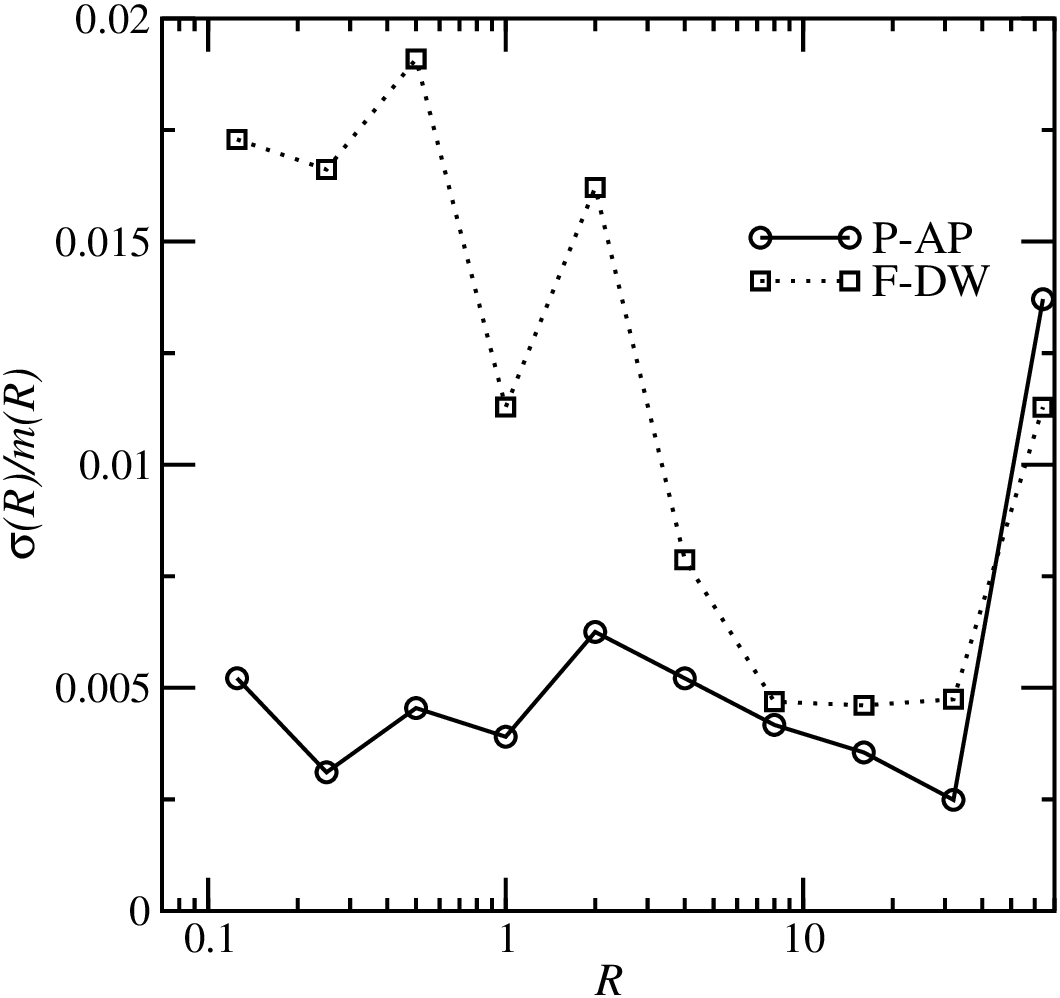}}
\end{center}
\caption{Standard deviation $\sigma(R)$ of scaling plot points devided by
  mean $m(R)$. The mean was taken over all considered values of $M$. 
Lines are guides to the eyes only. } 
\label{figScaleVarR}
\end{figure}

In Fig.\ \ref{figScaleVarR} the normalized standard deviation of the
scaling plot is shown as a function of $R$. Again, the high quality of
the scaling can be seen. Note that the statistics for both P-AP
and F-DW cases are comparable, while the statistics for $R=64$ are worse 
than for the other values due to the smaller number of samples. From 
Fig.\ \ref{figScaleVarR} we see that, for small values of $R$, the
F-DW case scales less well than the P-AP case. This may explain, why, for F-DW, the
stiffness exponent obtained for small aspect-ratios differs
significantly from the $R\to\infty$ limit. The fact that F-DW scales
less well than P-AP is probably due to the larger influence of the system boundaries
induced by the free bc. As a result, the scaling near $R=1$ is more complicated
than a simple $M^{1+\theta}$ behavior.

\begin{figure}[htb]
\begin{center}
\myscalebox{\includegraphics{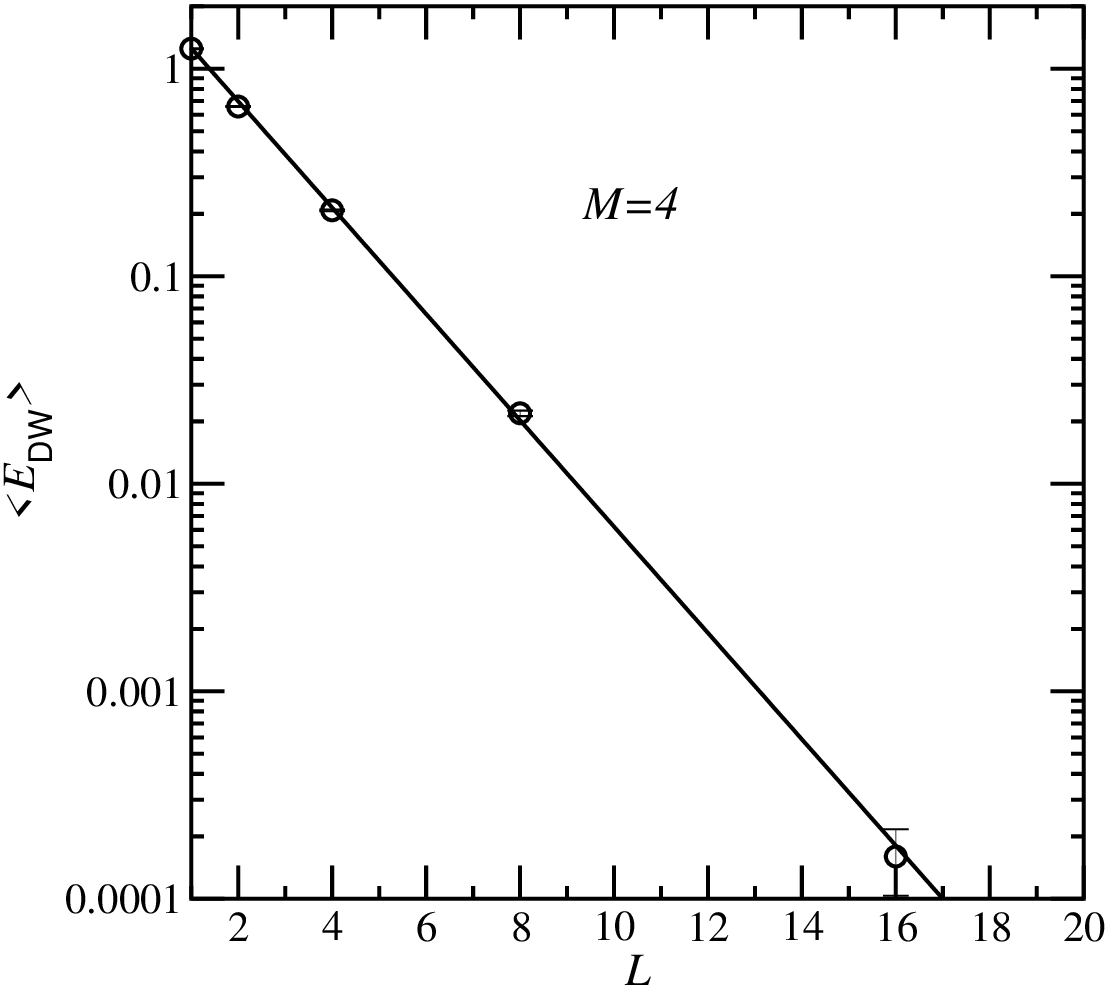}}
\end{center}
\caption{Domain-wall energy $\langle E_{\rm DW}\rangle$ of $\pm J$
  system with P-AP boundary conditions as a function of $L$ for $M=4$.
The line represents the function $g(L)=2.27\exp(-0.59L)$.}
\label{figPmf0_M4Stiff}
\end{figure}

Finally, we turn to the system with bimodal distribution of the
bonds. In Ref.~\onlinecite{stiff2d} a saturation of the domain-wall energy
has been observed for $R=1$ when $L=M\to\infty$, 
i.e.\ a fundamentally different result from
the Gaussian case. This raises the question of whether the
aspect-ratio approach is also able to make those results consistent.

As an example,  in Fig.\ \ref{figPmf0_M4Stiff} the domain-wall energy
is plotted for $M=4$ as a function of $L$. Clearly an exponential
decrease with $L$ can be observed. This is due to the discrete nature
of the interactions and can be explained as follows. For $L\gg M$,
the system can be decomposed into almost independent subsystems of size
$M\times M$. Let $p$ be the probability to find a {\em non-zero}
domain-wall energy in a subsystem. Then the probability to find a
non-zero $E_{\rm DW}$ in the full system is $p^{L/M}$ which decreases
exponentially with $L$.

As a result, for large $R$ (which implies a large value of $L$ for
fixed $M$), the probability to find a non-zero domain-wall energy is 
so small that for a reasonable number of samples $E_{\rm DW}$ will be 
exactly zero. This is what we have indeed found for $R=32$ and $R=64$. 
It means that for the bimodal model the aspect-ratio approach cannot 
be applied. This is in contrast to the Gaussian case, where aspect 
ratio scaling works better and better with increasing $R$.

\section{Summary and Discussion}

We have studied the scaling behavior of the energy of domain walls
in two-dimensional spin glasses 
obtained by changing the boundary conditions at $T=0$. For the
numerical calculations, we have employed a very efficient
matching algorithm, enabling us to calculate exact ground states of
systems up to size $N=1024\times 16$. 
To eliminate finite-size effects, we have applied
the aspect-ratio scaling approach.
Our main results are:
\begin{enumerate}
\item[(i)]
The value of the stiffness exponent for Gaussian distribution of
the bonds in $d=2$ seems to be
independent of the choice of the boundary conditions and the way the
domain walls are introduced. We give a final value of $\theta=-0.287(4)$.
\item[(ii)]
The scaling behavior predicted by the aspect-ratio approach 
applies very well for the Gaussian distribution of the bonds, 
again indicating that the value of $\theta$ is independent 
of the boundary conditions. 
\item[(iii)]
The aspect-ratio technique does {\em not} allow us to gain 
further insight for systems with the bimodal distribution of the bonds.
\end{enumerate}


\section{Acknowledgments}

The simulations were performed on a Beowulf Cluster at the 
Institut f\"ur Theoretische Physik of the Universit\"at
Magdeburg and at the Paderborn Center for Parallel Computing,  
both in Germany. The main part of the work was done during a 
visit of AKH to Manchester. Financial support for the visit was 
provided by the ESF programme SPHINX. AKH also obtained financial 
support from the DFG (Deutsche Forschungsgemeinschaft)
under grants Ha 3169/1-1 and Zi 209/6-1. ACC acknowledges support 
from EPSRC (UK). APY acknowledges support from the NSF through grant 
DMR 0086287 and the EPSRC under grant GR/R37869/01. He would also 
like to thank David Sherrington for hospitality during his stay at Oxford.

\end{document}